# Graphene-enabled Optoelectronics on Paper

Emre O. Polat, Hasan Burkay Uzlu, Osman Balci, Nurbek Kakenov,

Evgeniya Kovalska, Coskun Kocabas[†]

Bilkent University, Department of Physics, 06800, Ankara, Turkey

[†] Corresponding author email:

ckocabas@fen.bilkent.edu.tr,

Phone: +90 3122908078

**Abstract:** The realization of optoelectronic devices on paper has been an outstanding challenge due to the large surface roughness and incompatible nature of paper with optical materials. Here, we demonstrate a new class of optoelectronic devices on a piece of printing paper using graphene as an electrically reconfigurable optical medium. Our approach relies on electro-modulation of optical properties of multilayer graphene on paper via blocking the interband electronic transitions. The paper based devices yield high optical contrast in the visible spectrum with fast response speed. Pattering graphene into multiple pixels, folding paper into 3-dimensional shapes or printing coloured ink on paper substrates enable us to demonstrate novel optoelectronic devices which cannot be realized with wafer-based techniques.

The integration of semiconductors on unusual substrates enables devices with new mechanical functionalities that cannot be achieved with wafer-based technologies[1-8]. Electronic paper[9-12] has been the most attractive application aiming to reconfigure the displayed information electronically on a sheet of printing paper. Paper-based substrates have been an ambition of research in various fields ranging from medical diagnosis to display technologies aiming to create low cost and ubiquitous devices. For example, paper-based microfluidic devices[13], electronic circuits[14] even robotic systems[15] have been developed. Display devices on a sheet of printing paper has been a challenge[1, 9]. Several technologies based on electrophoretic motion of particles[11], thermochromic dye[10], electrowetting of liquids[12] and electrochromic materials[16] have been developed to realize electronic paper (e-paper) which has great potential for consumer electronics. Contrasting the primary aim of e-paper, these technologies, however are not compatible with conventional cellulose based printing papers. Here, using multilayer graphene as an electrically reconfigurable optical medium, we demonstrate an optoelectronic framework compatible with a conventional printing paper.

Optical properties of graphene can be controlled by doping. Doping alters the rate of interband and intraband electronic transitions of graphene and yields electro-modulation of optical absorbance in a very broad spectrum[17-19]. Optical contrast achieved by atomically thin graphene is limited by the optical absorption of 2.3 %, which is defined by fundamental constants[20]. In order to increase the optical modulation, graphene has been integrated on resonant optical cavities which lead to enhanced light-matter interactions via multiple passes[21]. In another approach, increasing number of graphene layers provide very large optical absorption which can be modulated by intercalation of ions into the layers which results in blocking of interband electronic transitions[22-25]. In our previous work, we showed that multilayer graphene films yield high-contrast optically reconfigurable medium which is suitable for display applications on unconventional substrates. In this letter, by integrating large area multilayer (ML) graphene on a piece of printing paper, we managed to fabricate optoelectronic devices on paper using electro-modulation of graphene layer via reversible intercalation process. The paper device consist of two multilayer graphene layers transfer-printed on both sides of the paper. In this configuration, ML-graphene simultaneously operates as the electrically reconfigurable optical medium and electrically conductive electrodes. In addition, the paper substrate yields a flexible and foldable mechanical support for the graphene layers and it holds the electrolyte (room temperature ionic liquid) in the network of hydrophilic cellulose fibres.

Figure 1 summarizes the optical and electrical properties of ML-graphene-on-paper. We synthesized ML-graphene on Ni foils using a chemical vapour deposition system using methane as carbon source. After the growth, we etched the metal substrate that leads a free standing graphene film on water surface. By immersing the paper substrate into the liquid we managed to transfer large area ML-graphene on printing paper (Supplementary Figs 1 and 2) . We used a piece of thick printing paper (Koehler Chamois Ivory Board, $246 g/m^2$) obtained from a local printing service. After the transfer process, we dried the paper at 70 °C and obtained conformal graphene coating on the network of fibres. Swelling of paper substrate during the transfer process and the surface roughness limits the quality of the MLG films. Figure 1a shows a photograph of ML-graphene on paper. Figure 1b shows a confocal microscope image of graphene coated surface. The RMS surface roughness of the paper is around 3 µm and the thickness is 300 µm (Figure 1c and Supplementary Figs 3-8). The scanning electron microscope image shows the rough surface consist of random network of fibres. The rough surface of paper provide an ideal Lambertian-view characteristics. Figure 1e shows the angular dependence of the scattering cross section of the paper surface (red line) and the Lambertian cosine model (black curve). The good agreement between the model and scattering data indicates the angle independent appearance of the paper surface. The graphene coated surface (Figure 1e, blue curve) yields more specular reflection than bare paper due to the metallic like nature. The characteristics of the diffuse reflectance varies from metallic-like to paper-like as the thickness of ML-graphene layer gets thinner (Supplementary Fig 9). At the same time, the sheet resistance of ML-graphene varies from 550 to 25 $\Omega$/sq as the number of graphene layers increases from 35 to 60 (Figure 1f). Besides it has an ideal-Lambertian view, the paper substrate yields a flexible and foldable support. Figure 1g shows the variation of resistance of ML-graphene layer after various acute folding scheme. We observed only 10% increase (Figure 1h) in the total resistance after acute folding which yields some structural defects on graphene (see the inset in Figure 1g). The folding introduce structural defects on graphene due to tearing apart of the paper surface and fold breaks on the surface. MLG films can be bended with much larger curvatures.

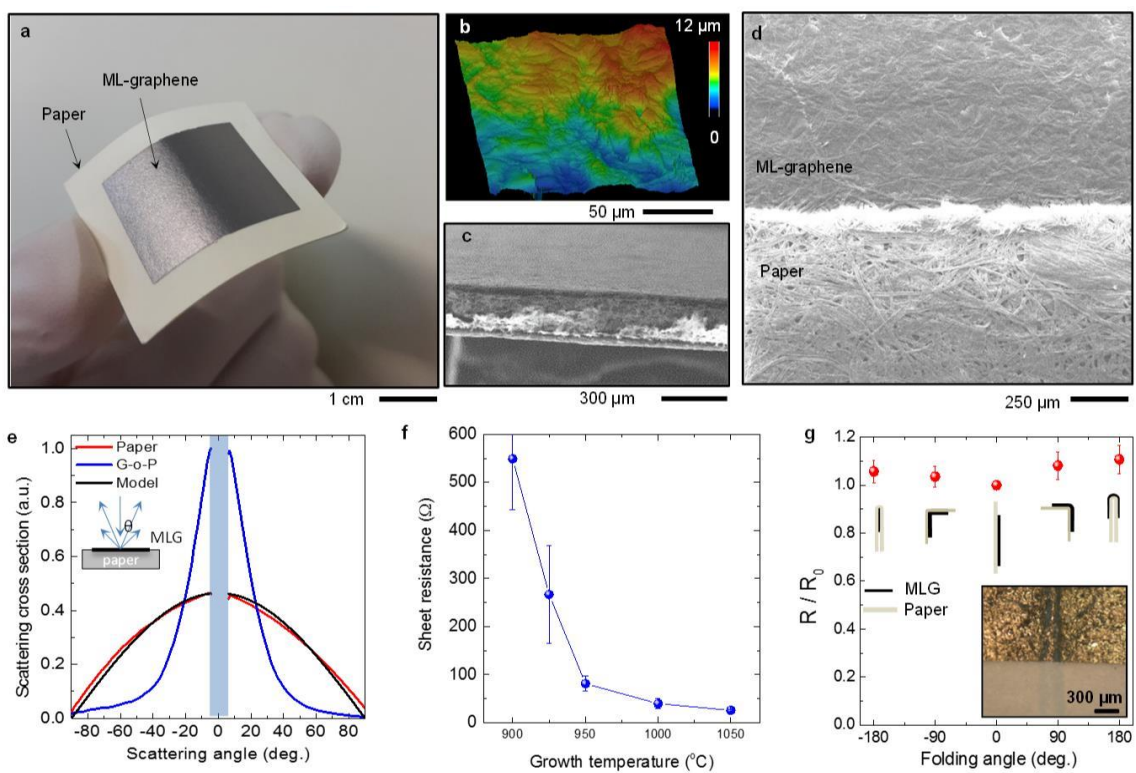

**Figure 1: Multilayer graphene on paper**: **a**, Photograph of multilayer graphene transferred on printing paper. **b**, Confocal microscope image of the graphene coated paper surface. The RMS surface roughness is around 3 µm **c,** Confocal microscope image of the rough paper surface coated with ML-graphene. **d,** Scanning electron microscope of the cross section and edge of graphene layer on paper. The bright area is due to electrostatic charging of the paper. The RMS surface roughness is around 3 µm. **e**, Scattering cross section at 650 nm of bare paper surface (red curve) and graphene coated surface (blue curve). The black curve shows the Lambertian cosine model. The inset shows the scattering configuration. **f**, The sheet resistance of ML-graphene-on-paper synthesized at different temperatures. **g**, Variation of resistance of ML-graphene-on-paper for various folding configurations. The inset shows the structural defects formed after folding.

Now, we would like to demonstrate electrically controlled colour change of ML-graphene-on-paper. Figure 2a shows the schematic drawing of the paper device. We transferred two ML-graphene layers on both side of the paper and soaked ionic liquid electrolyte into the paper. We used a silver-based conductive paint to make electrical contacts on graphene. Figure 2b and 2c show photographs of the fabricated device at different bias conditions. When we applied a bias voltage larger than 3 V, we observed a significant change in the colour. Figure 2d and 2f shows the corresponding optical microscope image of the edge of ML-graphene. Initially, ML-graphene has shiny metallic appearance, when we applied a voltage (> 3V) between the electrodes, it first gets dark and then become semi-transparent (See Movie 1, Figure 2d and 2e). To quantify the colour change, we measured the scattering cross section of graphene coated paper surface as we tuned the bias voltage from 0 to 4V (Figure 2f). At 0 V, device yield a significant specular reflection, however, as the voltage increases, the back-scattered intensity is suppressed resulting more Lambertian view due to the underneath paper surface. The colour of ML-graphene on paper can be switched between the metallic/black and semi-transparent view in a few second. Figure 2g and 2h show the time trace of the normalized intensity. Under a bias voltage, the anions of the ionic liquid intercalate into the graphene layers and block the interband transitions in the visible spectrum. The intercalation cycle takes relatively longer (~ 4 s) than the de-intercalation cycle (< 0.5 s). We also observed that the discharging current is significantly larger than the charging current due the high conductivity of the intercalated electrode (Supplementary Figs 10). The switching speed scales inversely with the area of the device. For a device with 3x3 mm$^2$ area, the response time is less than a second. The spectrum of the colour change covers nearly all visible spectrum (Supplementary Fig 11), however, the contrast diminishes for wavelength less than 500 nm.

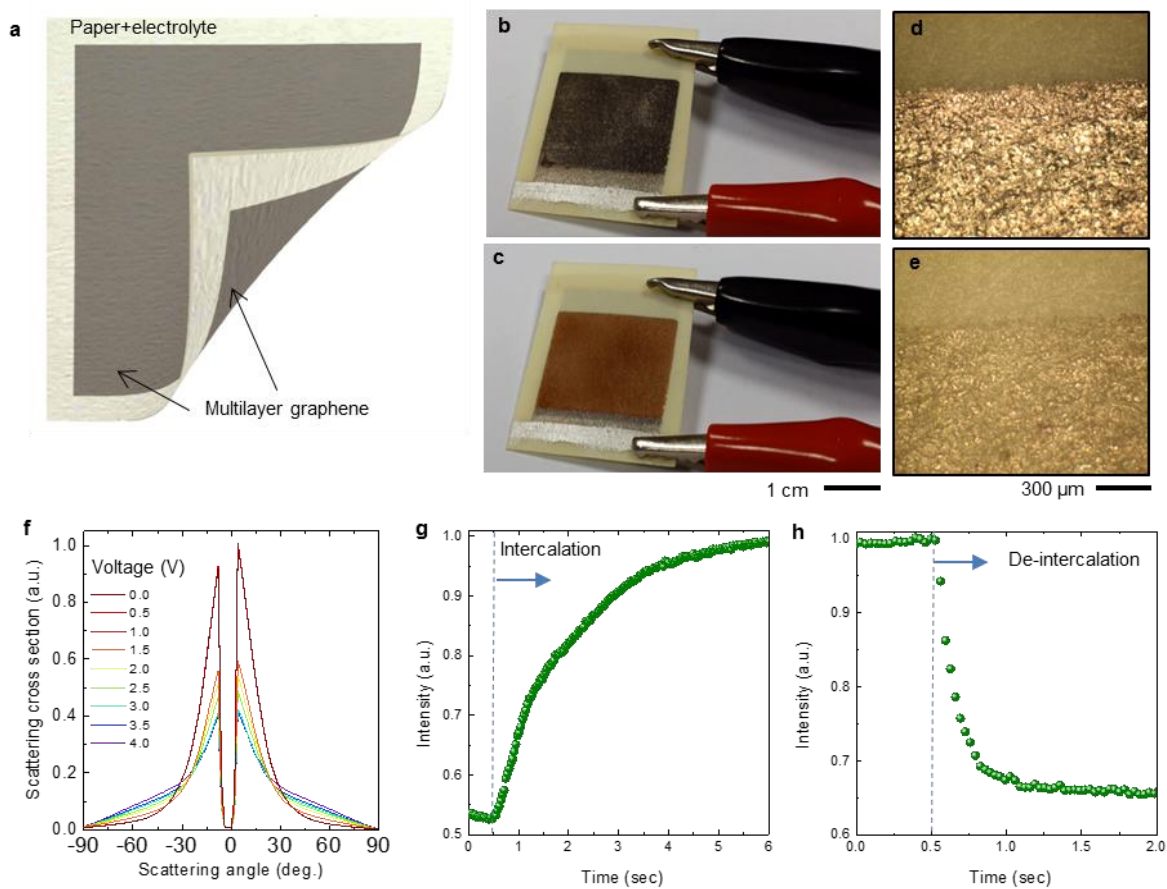

**Figure 2: Electrically controlled colour change on paper. a**, Schematic representation of the graphene based electronic paper consist of two multilayer graphene electrodes on paper substrates. The room temperature ionic liquid is soaked into the paper. **b,c** Photographs of the fabricated device under various bias voltage. **d,e**, Optical microscope images of the edge of ML-graphene at 0 and 4 V respectively **f**, Scattering cross section (at 600 nm wavelength) under bias voltages between 0 to 4 V. **g,h** Temporal behaviour of the colour change during intercalation and de-intercalation cycles, respectively.

The profound colour change is mainly because of blocking of interband transition of ML-graphene[22,26-27]. The weak interlayer coupling and very strong intralayer bonding of graphene sheets allow small atoms and molecules to intercalate between the layers[28]. In our case, under a voltage bias the anions of the ionic liquid intercalate into the layers and dope the graphene. As a result of intercalation, the charge density on graphene increases significantly and Fermi level shifts to higher energies about 1 eV. Unlike single layer graphene, ML-graphene has many conduction and valence subbands. The electronic transitions between these subbands result broadband absorption which can be suppressed by doping due to Pauli blocking[24]. To understand the mechanism of colour change on paper, we performed *in situ* electrical and optical measurements. Figure 3a shows the variation of the sheet resistance of graphene electrode during the intercalation cycle. The inset in Figure 3a shows the four-wire resistance measurement system. We slowly changed the bias voltage applied between the graphene electrodes and recorded variation of the sheet resistance. For undoped state, the sheet resistance is around 240 Ω/sq. As the bias voltage increases, the sheet resistance decreases substantially down to a value of 10 Ω/sq. This high conductivity of the intercalated ML-graphene confirms very high level of doping and charge densities. We observed that intercalation of acceptors (anions) in to the graphene is much efficient than the donors (cations). The bottom graphene electrode (connected to lower voltage) do not yield a visible colour change. We estimate that the shift in Fermi energy is around 1.2 eV, which yield a broad optical transmittance window. The window is defined by the cut-off of Pauli blocking ($2E_F$) and plasma frequency due to the free carriers[27, 29] ($\omega_p^2 = \frac{Ne^2}{m\varepsilon_\infty}$ where $N$ is the carrier concentration, $m$ is the effective mass, $e$ is the elementary charge and $\varepsilon$ is the dielectric constant). Increasing carrier concentration results a plasma frequency near the visible spectrum. As a result of these two opposite effects, transmittance spectrum shows a broad peak centred at the plasma frequency of the ML-graphene. Figure 3b shows the spectrum of the optical transmittance at different voltages (Supplementary Fig 12). The transmittance of the ML-graphene electrode reaches up to 90% at 820 nm. The energy of the transmittance maxima yields a plasma frequency of 1.5 eV. We also recorded *in situ* Raman spectra of the ML-graphene during intercalation (Figure 2f). We observed a significant enhancement and blue-shift in G-band (around 43 cm$^{-1}$), whereas the intensity of the 2D-band diminishes. The upshift in the frequency of G-band is due to the difference in the force constants arising from the intercalated environment. The upshift of 43 cm$^{-1}$ indicates acceptor intercalation of stage-1 which means anions intercalated into all layers. We do not observe a doublet structure which is a characteristic behaviour of partially intercalated graphitic compounds. The appearance of D-band demonstrate formation of defects during the intercalation process. These results reveal a very efficient and reversible intercalation process thus high level of doping of graphene layers. Our results suggest that ML-graphene-on-paper can be used as an electrically reconfigurable optical material which yields substantial colour change in the visible spectra.

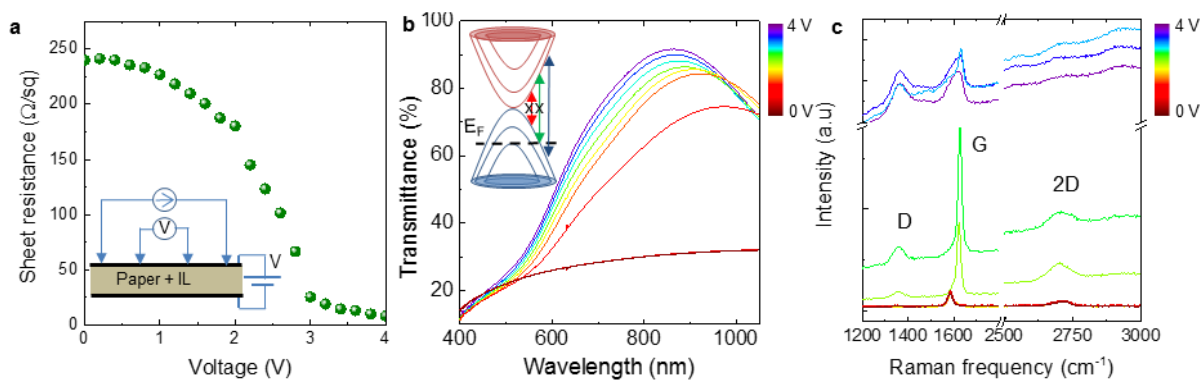

**Figure 3.** *In situ* **electrical and optical characterization of ML-graphene on paper: a**, Variation of the sheet resistance of ML-graphene during the intercalation process as the bias voltage between the graphene electrodes varies from 0 to 4 V. The inset shows the schematic representation of the four-point measurement system. **b**, Optical transmittance of ML-graphene at different bias voltages. **c**, *In situ* Raman spectra recorded for the ML-graphene during the intercalation of ionic liquid.

Cutting, folding and printing are common processes for paper but not for optical materials. Combining these scalable methods with our paper devices opens new possibilities for novel optoelectronic devices. To show the promises of this approach, we would like to demonstrate various device strategies. The first design is a planar display device consist of pattered MLG electrodes. We transferred MLG on one side of the paper and pattered the graphene electrode in to individually addressable shapes (i.e. numbers) using a computer controlled cutter. Figure 4a and 4b illustrate the layout and the photograph of the operating device with a back illumination (See Movie 2 and Supplementary Figs 13-15 ). The isolated elements and the rest work as the coplanar electrodes. The colour of each element can be switched under a voltage bias of 4V. We observed that the performance of the coplanar devices are similar with the devices with double electrodes. The device can operate even after we crumbled up the paper substrate (Supplementary Figs 16).

Folding a planar paper device into three dimensions enables new 3D display devices. Figure 4c shows the folding scheme for 3D paper display. We first pattered the number-shaped elements on the planar device then folded the substrate. Folding yields mountains and valleys on the surface and display elements on the planar faces. Figure 4d shows the picture of operation of the folded device. The graphene electrode preserve electrical continuity over the device even after many folding cycles. Furthermore, printing toner on paper allows us to add new functionality to our display devices. We demonstrated a colour principle by simply printing colour toner on the paper. Under a voltage bias, graphene electrode gets transparent and the printed colour appears. Figure 4e shows the schematic of the device. We printed a mosaic pattern using halftone colour. Halftone colours consisting of separated dots which allow ions to intercalate the graphene electrode. Under the voltage bias, graphene gets transparent and makes the colour pattern on the paper visible. Figure 4f shows the photograph of the device at 0 and 4V bias voltages. Red, yellow and green colours can be seen clearly trough the doped ML-graphene layer (Supplementary Figs 17). However, the blue dots on paper do not appear due to the inefficient optical modulation.

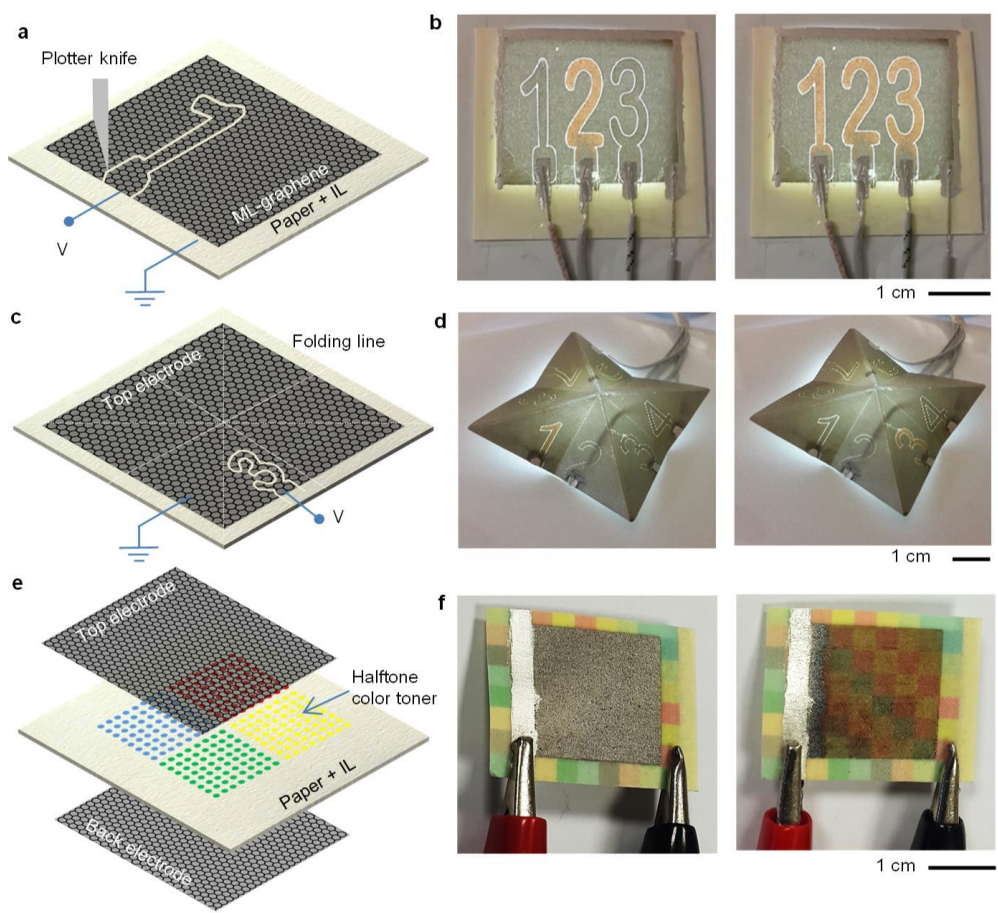

**Figure 4 Planar, 3D and coloured display devices on paper. a**, Schematic drawing of the planar paper display fabricated by patterning MLG electrode into isolated elements using a plotter. **b,** Photograph of the operating device. The colour of the isolated elements can be controlled individually. **c**, 3D paper display fabricated by folding the planar paper device into 3-dimension. The display elements are patterned on planar device. **d**, Photograph of the operating device with back illumination. The multilayer graphene electrode preserve its electrical continuity after the folding. **e**, Schematic drawing of the colour device formed by printing halftone colour toner on paper sandwiched between two ML-graphene electrodes. **e,f,** Photograph of the working device at 0 and 4 V bias voltage.

Now, we would like demonstrate a multipixel paper display. Figure 5a shows the schematic drawing of the 5x5 multipixel device formed by two patterned ML-graphene electrodes registered in perpendicular directions. Figure 5b shows the completed device with the wiring to the external control electronics. To address a pixel, we used passive matrix addressing by applying various bias voltage of 0, 2 and -2V to the rows and columns. When we applied 2 V to a column and -2 V to a row, the colour of the intersecting area changes from metallic to paper-view. Using a manual switch box, we were able to reconfigure the pattern by manually adjusting the voltage configuration. Figure 4b shows the images of the device with different reconfigurable pattern. (See Movie 3 for the real time operation). The migration of the boundary of the colour change illustrates the intercalation process. We were able to reconfigure the pattern on paper many time, however, we observed that after 30 minutes or voltages larger than 5V, the graphene layer gets oxidized and turn into dark back colour which cannot be revered. Degradation of the performance is likely due to the very reactive nature of highly doped graphene in air. The water content in ionic liquid is another detrimental factor on graphene. Undoped graphene is very stable in air however, highly doped graphene can be oxidize in ambient conditions. To reduce the oxidation, we first tested our device in vacuum. We observed that, we can apply more voltages (>5V) and the device can provide similar colour change. Coating the surface of graphene with a thin polymer layer improves the stability. (see Supplementary Figs 18) Our results suggest that the degradation of the device performance is due to oxygen in the atmosphere and water content in the electrolyte. The durability of the device increases when we bake the ionic liquid in vacuum before we inject it into the paper.

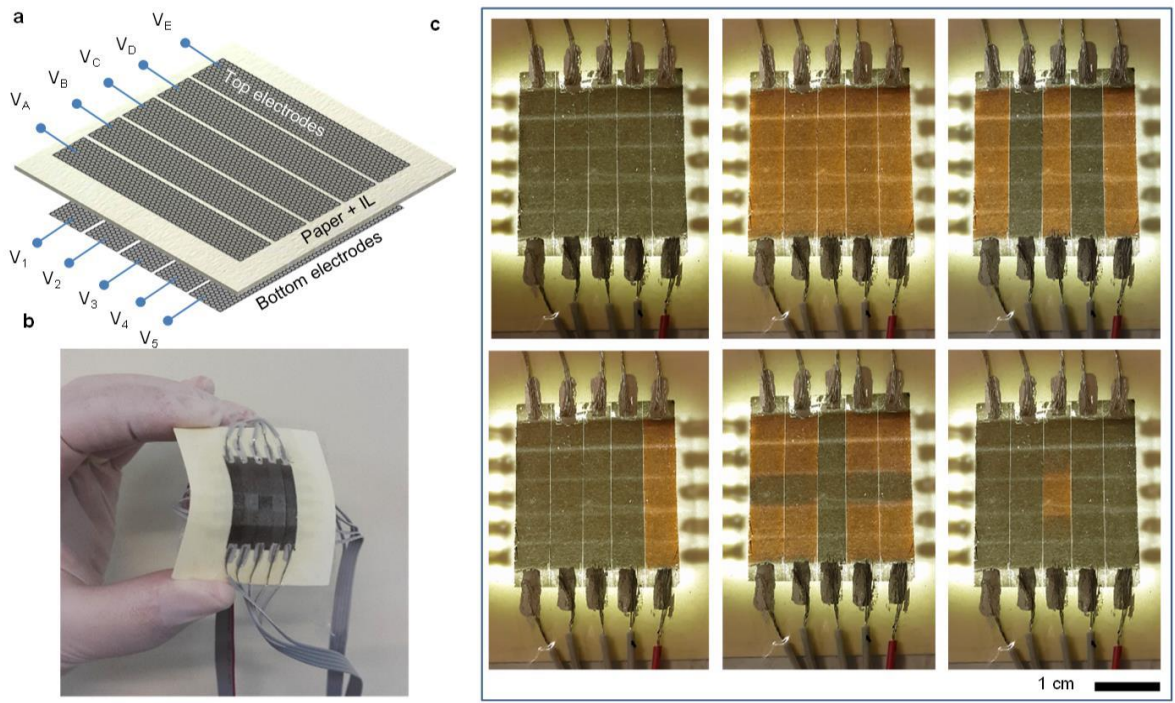

**Figure 5 Multipixel paper display**: a, Schematic representation of multipixel paper display using 5x5 array of patterned MLG electrodes transferred both side of the paper is a cross bar structure. The width and the length of graphene electrodes are 5 mm and 2.5 cm, respectively. **b**, Photograph of the fabricated paper display with the attached wiring. **c**, Photograph of the reconfigurable pattern on paper display using passive matrix addressing at various voltage configurations controlled by an electronic switch box.

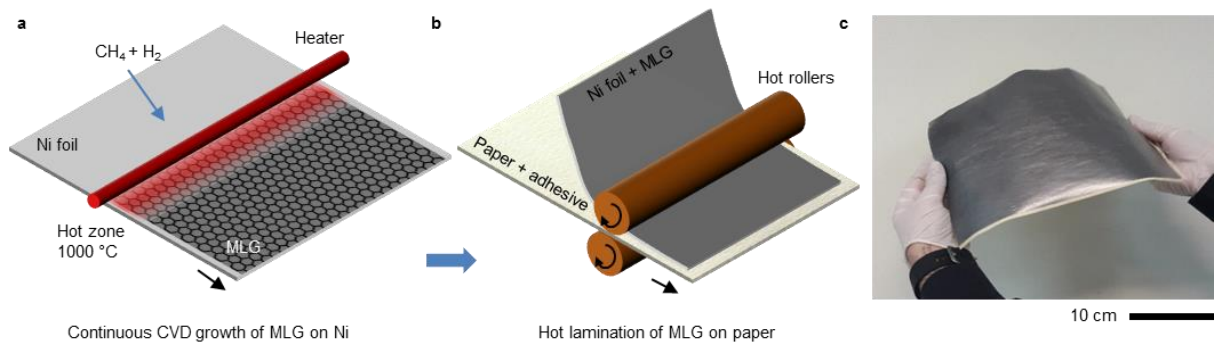

**Figure 6. Scaling up the process**. **a,** Schematic representation of the continuous chemical vapour deposition of multilayer graphene on Ni foils. **b,** Transfer process of multilayer graphene from Ni foil to paper substrate with hot lamination. The paper surface is partially coated with a polymer adhesive which holds the graphene layer and allows penetration of the electrolyte. **c,** Photograph of the multilayer graphene transferred on A4-size printing paper.

In the final section we would like to demonstrate compatibility of our method with roll-to-roll fabrication processes which play a key role for paper-based substrates. We modified our CVD system to handle large nickel foils. We grew multilayer graphene films at 1000°C on large area nickel foils by continuously moving the foil inside the chamber along the hot zone. Figure 6a summarises the continuous CVD process of MLG. Since the thickness of MLG is defined by the growth temperature rather than growth time, homogeneity of the temperature inside the chamber is a key parameter for the continuous growth. After the growth, we first tested the scalability of the transfer technique. With fishing technique, we were able to transfer MLG with size up to 6" in diagonal. Larger sample size results large cracks during the transfer process which prevents fishing methods for roll-to-roll process. Then we tested hot lamination based transfer processes. We partially coated the surface of paper with a very thin layer of thermal adhesive then we laminated MLG coated nickel foils on the paper surface. The partial coating is enough to hold the MLG film on paper surface and it allows penetration of ions. After etching the nickel foil, we obtained crack free large scale MLG films on paper (Figure 6c). The hot lamination technique and chemical etching is fully compatible with roll-to-roll processes.

We have shown that multilayer graphene on paper can be used as an electrically reconfigurable medium for display applications. In our device, graphene layers operates as both electrode and the optically active layer. The paper substrate provides flexible and foldable mechanical support which holds the electrolyte. Intercalation of multilayer graphene electrodes with room temperature ionic liquid yield high contrast reconfigurable colour change on a piece of printing paper. Scalable processing techniques used for paper manufacturing such as, cutting, folding or printing enable us to demonstrate novel paper based optoelectronic devices which cannot be realized with wafer-based techniques. We anticipate that our results provide a significant step for realization of low cost, disposable (Supplementary Fig 19) and ubiquitous optoelectronics on unconventional substrates.

**Methods Summary**

**Synthesis of multilayer graphene:** Multilayer graphene samples were synthesized on 50 µm thick nickel foil substrates (*Alfa Aesar item #12722*) using chemical vapour deposition system. The thickness of ML graphene samples were controlled by the growth temperatures varied between 900 $^0$C, to 1000 $^0$C. As the carbon feedstock, we used 30 sccm of $CH_4$ gas at the ambient pressure during the growth. To remove the oxide layer on Ni substrate, we anneled the samples under a flow of 100 sccm Ar and 100 sccm $H_2$ gases at $950^0$C. Growth time was 5 minutes then the samples were left for fast cooling to room temperature.

**Transfer-printing of multilayer graphene on paper:** Transfer printing of large area ML-graphene (up to size of 15 cm) on paper substrates were done by so called "fishing process". ML graphene is detached from the nickel substrate in dense $FeCl_3$ solution (1 M) and then it was transferred on clean water surface. Due to hydrophobic nature of graphene surface, MLG stays on the surface of water. Immersing the paper substrate into the liquid initiates conformal coating of the ML graphene on the paper surface. Finally the ML graphene holding paper was dried in the oven at 70 $^0$C for 2 hours to remove the water content which soaked during the transfer printing procedure.

**Preparation of the graphene-on-paper display devices:** After transfer printing of ML graphene to the both sides of the paper, we used silver based conductive ink on opposite sides of the top and bottom ML graphene as contact electrodes. When the conductive ink was dried 50 µL of ionic liquid electrolyte [DEME][TFSI] (98.5%, Diethylmethyl (2methoxyethyl)ammoniumbis(trifluoromethylsulfonyl)imide, Sigma-Aldrich, 727679) soaked into the paper . We attached two copper wires to apply voltage to the top and bottom ML graphene through the silver based contact electrodes.

**Scattering cross section measurements:** We measured the scattering cross-section from the bare and graphene coated paper surface using a supercontinuum laser (Koheras-SuperK Versa) integrated with an acousto-optic tunable filter. The wavelength can be tuned between 450 nm to 650 nm with 1nm spectral width. The incidence angle was controlled with a motorized rotary stage with an accuracy of 0.01 degree. The polarization dependent reflection from the metal surface is detected with a photodiode (Newport 818) connected to an amplifier. During the scattering measurements, we biased the graphene based paper displays by using Keithley 2400 source measure unit. The charging and discharging currents are also recorded simultaneously during the measurement.

**Raman spectroscopy**: We used Jobin Yvon Horiba Raman microscope system with a green laser having 532 nm excitation wavelength. The open device architecture of graphene based paper displays allowed us to measure the variations in the Raman spectrum of the ML- graphene under a bias voltage. The voltage was applied to the graphene based paper displays by Keithley 2400 source-measure-unit.

**In situ electrical measurements**: 4-point resistance measurement system is used to measure contact resistance of ML-graphene during intercalation and de-intercalation. We used two separate source meters (Keithley 2400 and 2600) to measure sheet resistance and apply bias voltage between graphene electrodes to initiate intercalation.

**Supporting Information**

Transferring ML-graphene on paper, variation of the resistance of ML-graphene under continuous folding, 3D-topography measured with an optical profiler, effect of surface roughness of paper on MLG films, transmittance and reflectance spectra of the ML-graphene on transparent PVC substrates, angular dependent normalized scattering cross section of paper coated with ML-graphene grown at different temperature, charging and discharging current intercalation and de-intercalation cycles, stability test of the devices in vacuum, disposable devices.

**Acknowledgements**

This work was supported by the Scientific and Technological Research Council of Turkey (TUBITAK) grant no. 113F278. C.K. also acknowledges the support from the European Research Council (ERC) Consolidator Grant ERC – 682723 SmartGraphene.